\documentclass[9pt,twocolumn,twoside]{pnas-new}
\templatetype{pnasresearcharticle} % Choose template 
\makeatletter
 \newcommand{\bc}{\begin{center}}
 \newcommand{\ec}{\end{center}}
                   \newcommand{\bfr}{\begin{flushright}}
                   \newcommand{\efr}{\end{flushright}}
   
     \newcommand{\be}{\begin{enumerate}}
     \newcommand{\ee}{\end{enumerate}}
        \newcommand{\bi}{\begin{itemize}}
        \newcommand{\ei}{\end{itemize}}
            \newcommand{\bd}{\begin{description}}
            \newcommand{\ed}{\end{description}}
                \newcommand{\beq}{\begin{equation}}
                \newcommand{\eeq}{\end{equation}}
                  \newcommand{\bea}{\begin{eqnarray}}
                  \newcommand{\eea}{\end{eqnarray}}

      \newcommand{\bfi}{\begin{figure}}
      \newcommand{\efi}{\end{figure}}
\newcommand{\bay}{\begin{array}{l}}
\newcommand{\eay}{\end{array}}
            
    \newcommand{\pa}{\partial}
    \newcommand{\del}{\delta}

    \newcommand{\sig}{\sigma}
    \newcommand{\eps}{\epsilon}
    \newcommand{\vf}{\varphi}

 \newcommand{\nnabla}{\mbox{\boldmath $\nabla$}}

\newcommand{\qq}{\mbox{\boldmath $q$}}
\newcommand{\KK}{\mbox{\boldmath $K$}}

\newcommand{\ff}{\mbox{\boldmath $f$}}

\makeatother

\usepackage{babel}

\begin{document}

\onecolumn
 
\thispagestyle{empty}
        \hspace*{1mm}  \vspace*{-0mm}
\noindent {\footnotesize {{\em
\hfill     \Large Posted on ArXiv, submitted to PNAS} }}
    %NOTE: Jjjj = journal name, abbreviated. Or write, Ԋjjj, in pressԍ
\vskip 1in
\begin{center}
{\Huge {\bf  Branching of Hydraulic Cracks in Gas or Oil Shale with
Closed Natural Fractures: How to Master Permeability }
           }\\[20mm]

{\huge {\sc Saeed Rahimi-Agham, Viet-Tuan Chau, Huynjin Lee, Hoang Nguyen, Weixin Li, Satish Karra,  Esteban Rougier, Hari Viswanathan, Gowri Srinivasan, Zden\v ek P. Ba\v zant}}
\\[1in]

{\Large \sf SPREE Report No. 18-10/076b}\\[1.5in]
%Here 33788 refers to my expired DoE Grant on ASR. ZPB
   %NOTE: 778 is the abbreviated project number; 09 stands for the
   %month, x stands for the first letter of the title, or other
   %if there is a confusion.
   
{\Large Center for Science and Protection of Engineered Environments (SPREE)
\\ Department of Civil and Environmental Engineering
%McCormick School of Engineering and Applied Science
%\\ Departments of Civil Engineering and Materials Science??
\\ Northwestern University
\\ Evanston, Illinois 60208, USA
\\[1in]  {\bf November 20, 2018}} %\\ Revised ... 2003
\end{center}
   %NOTE: Adjust spacing to fit one page, if needed

\clearpage   \pagestyle{plain} \setcounter{page}{1}

\twocolumn

\title{Branching of Hydraulic Cracks in Gas or Oil Shale with
Closed Natural Fractures: How to Master Permeability}

% Use letters for affiliations, numbers to show equal authorship (if applicable) and to indicate the corresponding author
\author[a]{ Saeed Rahimi-Aghdam}
\author[b]{Viet-Tuan Chau}
\author[a]{Huynjin Lee} 
\author[a]{Hoang Nguyen}
\author[a]{Weixin Li}
\author[b]{Satish Karra} 
\author[b]{Esteban Rougier} 
\author[b]{Hari Viswanathan} 
\author[c]{Gowri Srinivasan}
\author[d,1]{Zden\v ek P. Ba\v zant}

\affil[a]{Northwestern University, Department of Civil and Environmental Engineering, 2145 Sheridan road, Evanston, Illinois 60208}

\affil[b]{ Earth and Environmental Sciences Division, Los Alamos National Laboratory, Los Alamos, NM 87545, USA}

\affil[c]{X Computational Physics Division, Los Alamos National Laboratory, Los Alamos, NM 87545, USA }

\affil[d]{McCormick Institute Professor and W.P. Murphy Professor of Civil and Mechanical Engineering and Materials Science, Northwestern University, 2145 Sheridan road, Evanston, Illinois 60208}
%\affil[c]{Affiliation Three}

% Please give the surname of the lead author for the running footer
\leadauthor{Rahimi-Aghdam} 

% Please add here a significance statement to explain the relevance of your work
\significancestatement{Development of a realistic model of fracking would allow better control. It should make it possible to optimize various parameters such as the history of pumping, its rate or cycles, changes of viscosity, etc. This could lead to an increase of the percentage of gas extraction from the deep shale strata, which currently stands at about 5\% and rarely exceeds 15\%.}    %The benefits could be enormous.}

% Please include corresponding author, author contribution and author declaration information
\authorcontributions{Rahimi-Aghdam did the simulations and wrote the first draft, Ba\v zant designed the research and edited the draft. All the other coauthors contributed by discussions. }
%\authordeclaration{Please declare any conflict of interest here.}
%\equalauthors{\textsuperscript{1}A.O.(Author One) and A.T. (Author Two) contributed equally to this work (remove if not applicable).}
\correspondingauthor{\textsuperscript{2}To whom correspondence should be addressed. E-mail: z-bazant@northwestern.edu}

% Keywords are not mandatory, but authors are strongly encouraged to provide them. If provided, please include two to five keywords, separated by the pipe symbol, e.g:
    % $|$ Fracture Mechanics}
 %10/26 It makes no sense to duplicate words frm the title because they are caught by computer search anyway.

\begin{abstract}
While the hydraulic fracturing technology, aka fracking (or fraccing, frac), has become highly developed and astonishingly successful, a consistent formulation of the associated fracture mechanics that would not conflict with some observations is still unavailable. It is attempted here. Classical fracture mechanics, as well as the current commercial softwares, predict vertical cracks to propagate without branching from the perforations of the horizontal well casing, which are typically spaced at 10 m or more. However, to explain the gas production rate at the wellhead, the crack spacing would have to be only about 0.1 m, which would increase the overall gas permeability of shale mass about 10,000$\times$. This permeability increase has generally been attributed to a preexisting system of orthogonal natural cracks, whose spacing is about 0.1 m. But their average age is about 100 million years, and a recent analysis indicated that these cracks must have been completely closed by secondary creep of shale in less than a million years. Here it is considered that the tectonic events that produced the natural cracks in shale must have also created weak layers with nano- or micro-cracking damage. It is numerically demonstrated that a greatly enhanced permeability along the weak layers, with a greatly increased transverse Biot coefficient, must cause the fracking to engender lateral branching and the opening of hydraulic cracks along the weak layers, even if these cracks are initially almost closed. A finite element crack band model, based on recently developed anisotropic spherocylindrical microplane constitutive law, demonstrates these findings.
\end{abstract}

\dates{This manuscript was compiled on \today}
\doi{\url{www.pnas.org/cgi/doi/10.1073/pnas.XXXXXXXXXX}}

\maketitle
\thispagestyle{firststyle}
\ifthenelse{\boolean{shortarticle}}{\ifthenelse{\boolean{singlecolumn}}{\abscontentformatted}{\abscontent}}{}

\dropcap{S}ignificant advances have been made in fracture mechanics of propagation of a single hydraulic crack in elastic rock under tectonic stress  \cite{Abe76,Bec10,Ada08,Adv87,Car97,Det03,Ada02,Val95,Zob10,Det16,Det04}. They include characterization of the stress singularity at the tip of a water-filled advancing crack, flow of water of controlled viscosity along the crack, with or without proppant grains, and water leak-off into the shale. Interactions of parallel cracks, their stability, closing, and stress-shadow effect, have also been clarified \cite{Baz85,Baz14,Bun05,Bung15}. Discrete element models used in most commercial softwares, in which the hydraulic crack was simulated by a band of inter-element separations \cite{FracCADE,StimPlan}, led to similar results.

These studies, however, predicted no branching of the hydraulic cracks,  originally spaced at cca 10 m. This  presented a dilemma since branching is the only way to reduce the crack spacing to about 0.1 m, which is necessary to explain the gas production rate. Consequently, it has been universally hypothesized that the preexisting natural cracks, spaced at cca 0.1 m,
    %produced in past tectonic events 
would somehow increase the overall permeability of the shale mass. A %An enormous 
10,000-fold increase of permeability would be necessary to match the gas production rate. But recent analysis \cite{Chau17,BazR17} showed that the natural, tectonically produced, cracks, which are on the average about 100 million years old, must have been closed by secondary creep (or viscous flow) of shale under tectonic stress within 10,000 to 1 million years (if not filled earlier by calcite deposit). This invalidated the hypothesis.

It might be objected that water in the cracks could have prevented crack closing. But the open spaces in shear cracks, created (due to shear dilatancy) by a tectonic event, could not have been filled by water immediately. If the water had to seep in from the ground surface, it would take about 10 million years, if from a nearby water-filled rock formation, certainly over a million years. This must have left plenty of time for the creep closing to proceed uninhibited.

\begin{figure}[h]
\begin{center}
\includegraphics[width=0.67\linewidth,height=0.7\linewidth]{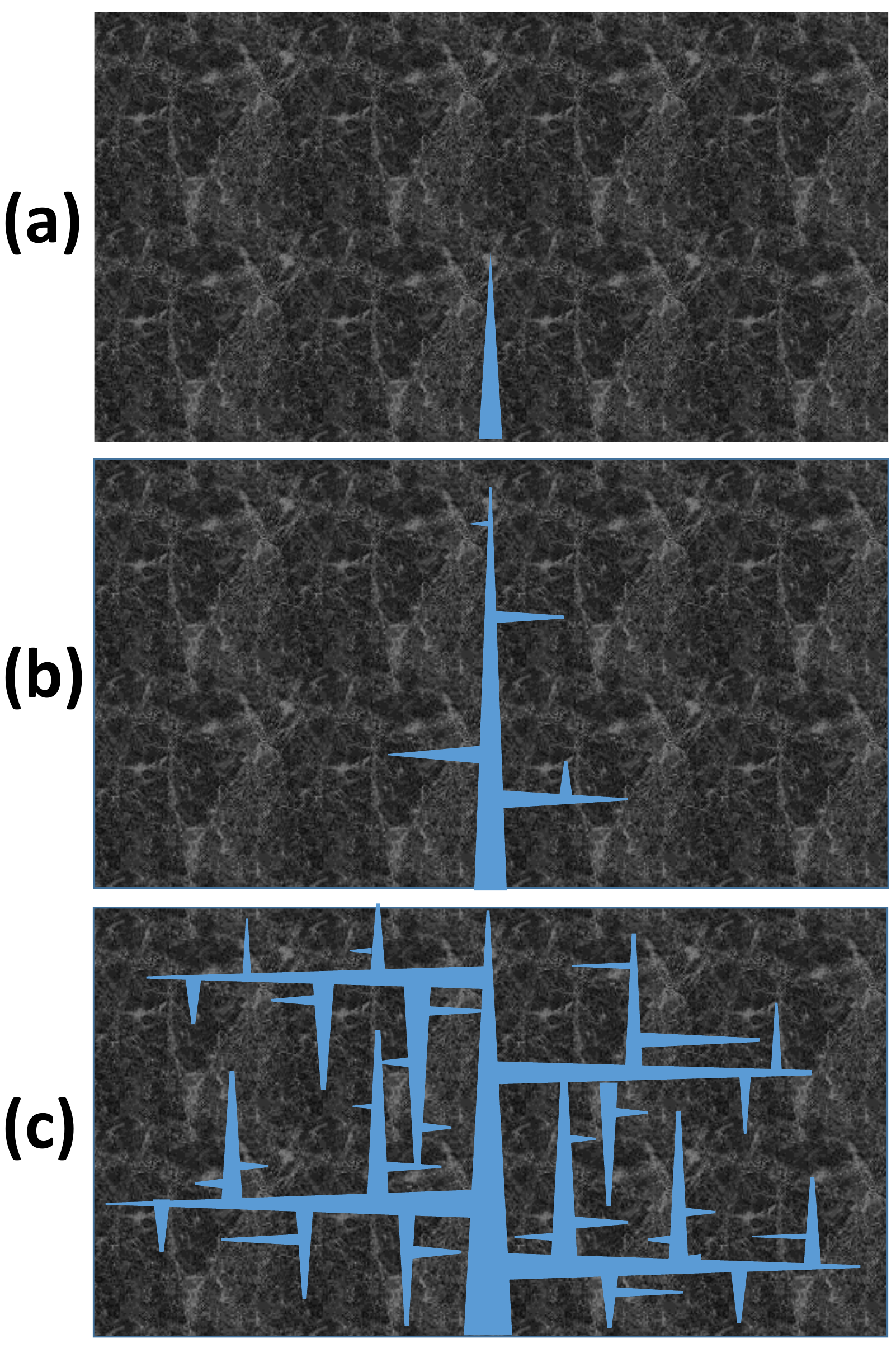}
\caption{Schematic Schematic branching due to natural fractures; a) water is injected at high pressure through damaged zones and weak layers, b) crack branching initiates due to the presence of damaged zones and natural fractures, and c) dense cracking happens in all directions due to the presence of damaged zones, weak layers at closed natural fractures (downward view normal to bedding plane) \label{fig:Frack}}
\end{center}
\vspace{-0.4cm}
\end{figure}

A recent paper \cite{Chau16} presented a new model which, by contrast with all the previous studies, took into account: 1) the seepage forces (i.e., the body forces due gradients of pore pressure in Darcy diffusion of water into porous shale), and 2) the variation of effective Biot coefficient for the water pressure on the crack plane, caused by gradually vanishing bridges between the opposite faces of a widening bridged crack (another difference from the previous studies was to abandon the assumption of incompressibility of water in the cracks, since water is about 20-times more compressible than shale). This model \cite{Chau16} did predict extensive lateral crack branching.

Later analysis, however, showed that the branching indicated by the computer program in \cite{Chau16} was, in fact, triggered by the unintended coding of a sudden change of Biot coefficient for transverse water pressure on the crack. This change abruptly increased the water pressure on the solid phase and triggered dynamic response. Such a sudden trigger is probably unrealistic, which represents a vital correction to the preceding study \cite{Chau16} (this correction nevertheless reveals a useful fact, namely that producing shocks in fluid pressure could greatly enhance crack branching).

If the change in Biot coefficient is made gradual, the model from the previous study \cite{Chau16} would predict no crack branching, although the branching must occur to explain the observed gas production rate. This study will show that if the previous model \cite{Chau16} is enhanced by introducing, into the shale mass, significant heterogeneity due to damaged weak layers along preexisting natural cracks, then an extensive and dense crack branching is predicted.

%Saeed, revised by ZPB 10/26
It may be noted that the fracking companies are aware of the necessity of branched cracks running along preexisting natural fractures. Fig. \ref{fig:Frack} shows a picture similar to what is found on the websites of some companies. However, this awareness seems to be merely intuitive and empirical. The existing commercial softwares, as well as fracture mechanics studies, predict no branching. So an intersecting system of open natural fractures is assumed to either exist a priori or to develop according to some empirical criteria with no basis in mechanics, supported by some recent experiments indicating the possibility of branching \cite{Weng11,Blan82,Ols09,zeng17}. A physics-based model for branching, which is our goal, seems lacking.
      %It should be noted that big fracking companies are aware of crack branching due to the effect of natural fractures. Fig. \ref{fig:Frack} illustrates a common figure that can be found in the website of big oil companies to explain the fracking process. However, this knowledge seems mostly empirical and the current models for simulating the fracking process neglect the branching phenomenon and consider natural fractures to be open from beginning. Recently, several studies analyzed, experimentally and computationally, the possibility of crack branching due to the presence of natural fractures and the experimental results clearly show the possibility of branching. However, the current computational models in literature for simulating crack branching, either consider natural fractures to be open from beginning or just provide some empirical relations for the initiation of crack branching. Therefore, it seems a physics-based model that can simulate the crack branching due to closed natural fractures and weak layers seems lacking.

\subsubsection*{Fluid flow in porous solid, without or with cracking damage}

Two types of flow play a role in hydraulic fracturing: 1) The flow along the hydraulically created cracks, typically a few millimeters wide, and 2) the flow through nano-scale pores and micro- or nano-cracks in shale with preexisting damage. The latter is negligible after continuous hydraulic cracks form,
  %and has been disregarded in apparently all studies except \cite{ChaBazSu16}. But it dominates before cracks form
but here it is found to be crucially important for crack initiation and branching. The volume flow, $q$, of water through the pores and nano- or micro-cracks of isotropic material may be approximately calculated from the Darcy law:
 $ \qq = - (K / \mu) \nnabla \psi
 $       % 10/26 here K is not a matrix. ZPB
where $K$ = permeability, $\mu=$dynamic viscosity and $\psi =$ phase potential calculated as,
 $  \psi = p - \gamma_g z
 $
Here $p=$pore pressure, $\gamma_g = $pressure gradient due to gravity, and $z = $depth from a datum. However, the permeability $K_v$ in the direction normal to the bedding planes $(x,y)$, i.e., in vertical direction $z$) is much lower than permeability $K_h$ along these planes (horizontal). Therefore, the three-dimensional Darcy law is, in general, anisotropic. In Cartesian coordinates ($x,y,z)$, the resulting volume flux vector $\qq = (q_x, q_y, q_z)$ may be written as
 \begin{equation} \label{qq}
  \qq = - \mu^{-1} \KK \cdot \nnabla \psi
 \end{equation}
where $\nnabla$ is the vector of gradient operator; and $\KK$ is the $3 \times 3$ permeability matrix, which is diagonal if (and only if) the cartesian axes $x,y,z$ are chosen to be parallel and normal to the bedding planes.   %, which are the principal directions of permeability.
% traditionally for shale $K_{I}$ is denoted as %$K_{v}$ and $K_{II} = K_{III} = K_{h}$.

Although the natural (or preexisting) cracks in shale strata at 3 km depth must have been closed by hundred million years of creep, the damage bands along these cracks, which always accompany propagation of fracture process zone, certainly remain (in fact, based on the known surface energy of shale, it can be shown that even empty pores and cracks cca $<$ 15 nm in size, at depth 3 km, cannot close, and this is confirmed by the known size of pores containing shale gas).  Permeability $K_{xx}$ along these bands is surely much higher than it is in the intact shale 
% The effect of this difference will be calculated.
(but pores $<$15 nm contribute nothing globally). 

To prevent the formation of horizontal cracks, the pumping pressure is assumed not to exceed the overburden pressure, which is about 75 MPa. The hydraulic fracturing is considered to produce a system of mutually orthogonal vertical cracks, normal to the directions of the minimum and maximum principal tectonic stresses. The flow of the second type, along the hydraulically created cracks, may be assumed to follow the Reynolds equations of the classical lubrication theory, which are based on the Poiseuille law for viscous flow. Thus the horizontal and vertical flow vector components in $x, y, z$ directions along with the cracks may be calculated as
 \beq \label{Q}
  Q_x = - \frac{h_y^2}{12 \mu}\ \nabla_x p,~~~
  Q_y = - \frac{h_x^2}{12 \mu}\ \nabla_y p,~~~
  Q_z = - \frac{h_x^2 + h_y^2}{12 \mu}\ \nabla_z p
 \eeq
where $\nabla_x = \pa / \pa_x$,...; $h_x, h_y$ = opening widths of vertical cracks normal to axes $x$ and $y$ that are positioned into the bedding plane. 

\begin{figure}
\includegraphics[width=1\linewidth]{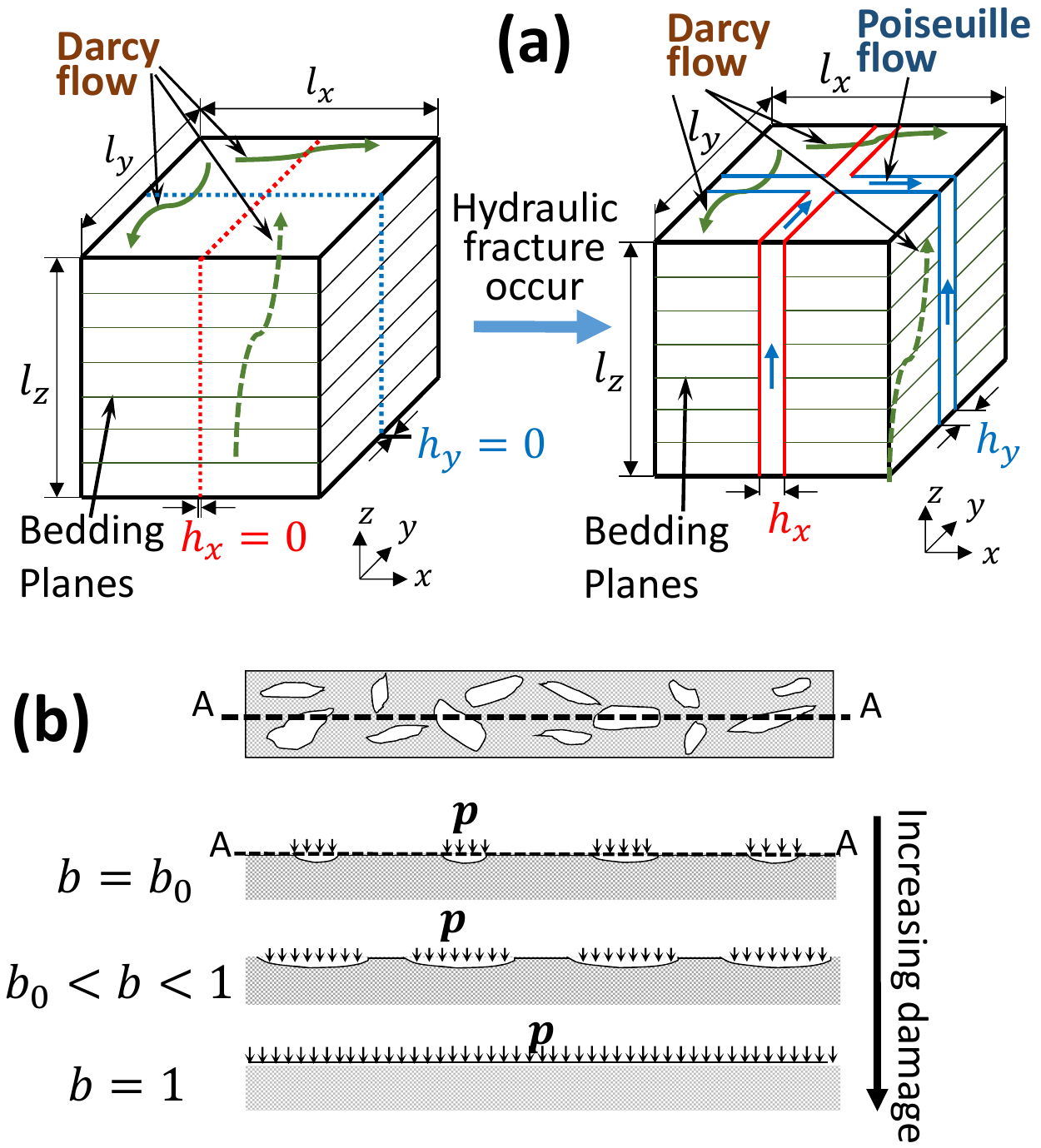}
\caption{a) Fluid flow in intact and damaged shale; b) Biot coefficient increase due to damage \cite{Chau16} . \label{fig:flow}}
\vspace{-0.4cm}
\end{figure}

An effective way to simulate the hydraulic cracks numerically is the crack band model \cite{Baz83,Cer05,Baz97}, in which cracking deformation is considered meared over the band (or element) width. The widths of cracks normal to $x$ and $y$ are: 
 \begin{equation}    \label{hxhy}
  h_x = l_x \epsilon"_{xx},~~~h_y = l_y \epsilon"_{yy}
 \end{equation}
(Fig.\ref{fig:flow}) where $\epsilon"_{xx}, \epsilon"_{yy}$ = damage parts of normal strains due to smeared cracking normal to $x$ and $y$ directions; $l_x, l_y$ = crack band widths, assumed equal to the minimum possible spacing of adjacent parallel hydraulic cracks ($l_x, l_y$ must be treated as a material property, related to fracture energy $G_f$ of shale; here $l_x, l_y$ are not changed but if they were the postpeak would have to be adjusted to preserve $G_f$). Furthermore,
 \begin{equation} \label{eps"}
  \epsilon"_{ij} = \epsilon_{ij} - \epsilon_{ij}^{el},~~~
  \epsilon_{ij}^{el} = C_{ijkl} \sigma_{kl}
 \end{equation}
where $C_{ijkl}$ = transversely isotropic elastic compliance tensor of shale (for unloading); $\sig_{ij}, \eps_{ij}$ are the stress and strain tensors in the rock, calculated from a constitutive model for smeared cracking damage (with a localization limiter \cite{Baz83}), for which the spherocylindrical microplane constitutive model \cite{Li17} has been used. The coordinates are Cartesian, $x_i, i=1,2,3$ ($x_1 \equiv x, x_2 \equiv y, x_3 \equiv z$).
    %Saeed 
Note that, the same as in \cite{Chau16}, water is  considered as compressible. It is, in fact, about 20-times more compressible than concrete, and the water pressure during fracking can be high (up to 70 MPa). 
          %In this study, for considering water compressibility, the same relations and values as \cite{Chau16} were used.

\subsubsection*{Equilibrium in two-phase solid and Biot coefficient}

The shale may be modeled as a two-phase medium with water-saturated pores, for which the classical Biot-type relations for the equilibrium of the phases apply. For undamaged shale, they read:
 \begin{equation} \label{Sij}
   S_{ij} = \sigma_{ij} - \delta_{ij} b_0 p
 \end{equation}
where $p=$ pore pressure, $b_0 =$ Biot coefficient of undamaged shale; $S_{ij}$ = total stress tensor; $\sig_{ij}$ = stress tensor in the solid phase, and $\del_{ij}$ = Kronecker delta. As a special case, $S_V = \sigma_V - b_0 p$ where $S_V = S_{kk}/3$ = volumetric total stress, and $\sig_{ij} = \sig{kk}/3$ = volumetric stress in the solid phase.

While typically $\vf = 0.1$, the Biot coefficient of shales can vary between $0.2$ and $0.7$.
       %and to depend on the ratio of the strengths
       %10/1 Strength? Isn't it stiffness? ZPB
Test results \cite{Hu10,Shoj14,Con04,Xie12,Shao98,Baz16} show that it increases with the cracking damage and depends on the load direction. This requires generalizing the Biot coefficient as a tensor, $b_{ij}$ \cite{ort07}. The following, tensorially consistent, empirical relation, which appears to match test data, is proposed:
 \begin{equation}  \label{bij}
  b_{ij} = \min \left\{ b_{0} + \beta \epsilon"_{ij} (\eps" _{kk}/3)^{-2/3},  \  1 \right\}~~~~(\vf \le b_0 \le 1)
 \end{equation}
Here $b_{0}$ refers to undamaged material, $\beta$ = empirical parameter, $\epsilon_{ij}''$ is the inelastic damage strain tensor, and $\vf$ = natural porosity of shale. For the Biot coefficient in the direction of unit vector $\nu_i$, this equation gives $b_\nu = \nu_i \nu_j b_{ij} = b_0 + \beta \eps"_\nu (\eps"_V)^{-2/3}$ (but $\le 1$), where $\eps"_V = \eps"_{kk} /3$ = inelastic relative volume expansion, and $\eps"_\nu = \nu_i \nu_j \eps"_{ij}$ = inelastic normal strain component in direction of vector $\nu_i$.

For the special case of micro- or nano-cracking normal to $x_1$ direction only, one has $\eps"_V = \eps"_{11}/3$ and $b_\nu = b_0 + \beta (9 \eps"_{11})^{1/3}$ (but $\le 1$). This equation can be interpreted graphically as seen in Fig. \ref{fig:flow}b, which shows section A-A of a band of preexisting, mostly aligned, microcracks and the compressive stresses applied by the pore fluid onto the microcrack faces, resisted by tensile stresses in the ligaments of the solid between the microcrack tips.

The viscous drag of water flowing through a soil imposes a seepage force on the soil in the direction of flow. The seepage forces are body forces defined as
 \beq \label{fseep}
  \ff_s\ = \ b \nnabla p
 \eeq
They are applied on the porous solid and must be balanced by stresses in the solid. Seepage in an upward direction reduces the effective stress within the soil. When the water pressure at a point in the soil is equal to the total vertical stress at that point, the effective stress is zero and the soil has no frictional resistance to deformation \cite{TerzBook,HotBook}. They have long been considered in geotechnical engineering to assess the risk of sand liquefaction in cofferdams \cite{Baz40,Baz53} or under dams. However (except for \cite{Chau16}) they have been ignored in previous studies of hydraulic fracturing, although they do play a crucial role in crack branching. A poromechanical finite element (FE) code for a two-phase solid automatically takes the seepage forces into account in the form of nodal forces.

\subsubsection*{Two-phase finite element (FE) simulations for a {\em single} damage band}

\begin{figure}[h]
\begin{center}
\includegraphics[width=1\linewidth]{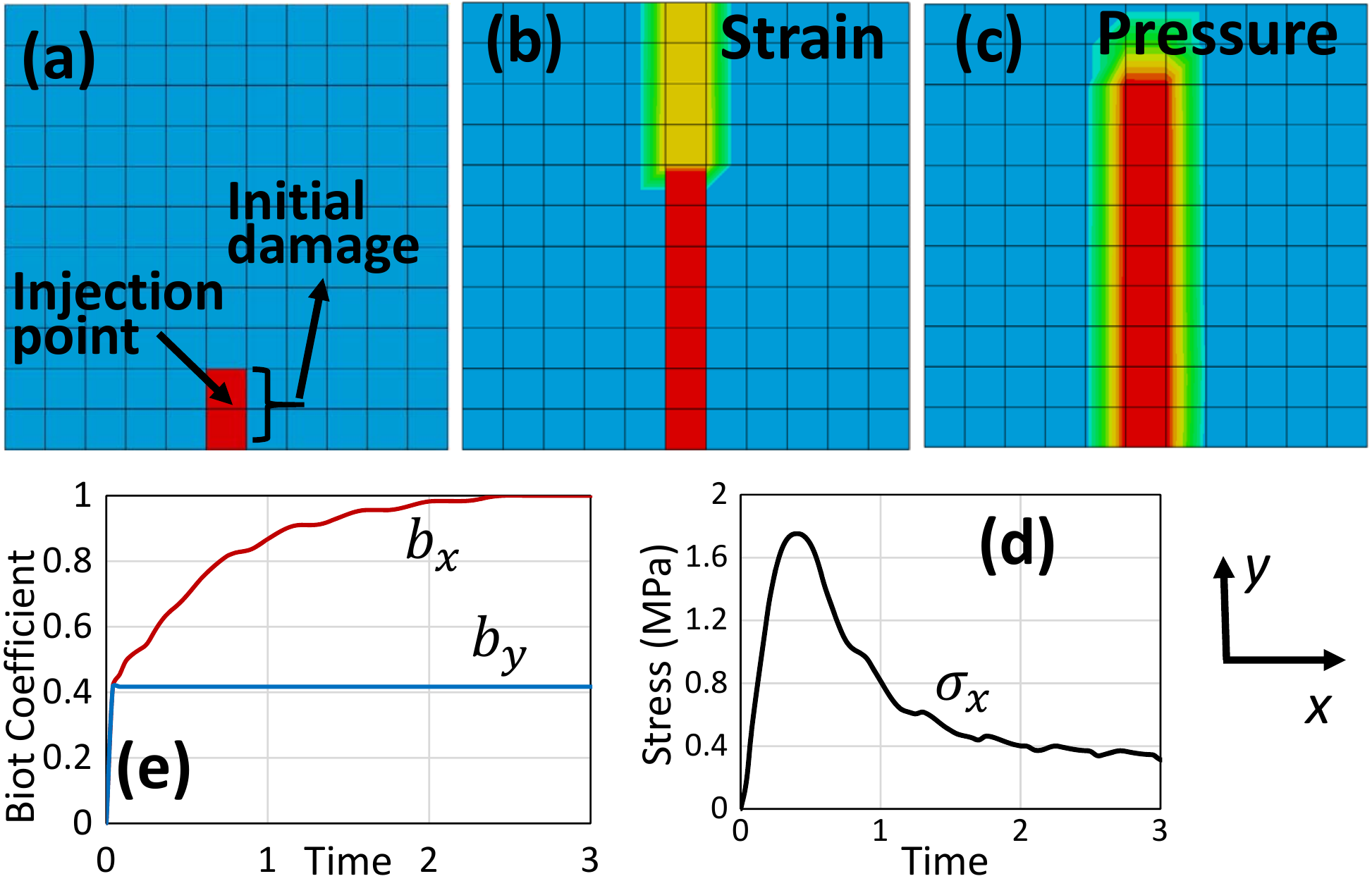}
\caption{Results of two-phase FE simulations in case of a {\em single} damage band. a) Initial model. b) Strain $\eps_{xx}$ due to water injection (the red  marks the highest values, the blue the smallest). c) Pressure propagation. d) Stress evolution vs. injection time for the first element above the initially damaged elements. e) Evolution of Biot coefficient transverse to, and along, the damage band.     \label{fig:verf}}
\end{center}
\vspace{-0.4cm}

\end{figure}

To clarify the role of nano- or micro-cracking, consider first a horizontal two-dimensional (2D) square block of shale of dimensions 1.1 m $\times$ 1.1 m, supported at the sides by springs approximately equivalent to an infinite medium, as shown in Fig. \ref{fig:verf}a. Water is injected at the center of south side at the constant rate of 2 m$^3$/s. The anisotropic spherocylindrical microplane model, with the default parameters of shale given in \cite{Li17}, is used as the constitutive model; and $l_x = l_y$ = 2.1 m. The initial Biot coefficient is $b_0 = 0.4$. The tectonic stresses are $T_x = - 30$ MPa, $T_y = - 30$ MPa.

Consider that there is a {\em single} preexisting band of nano- or micro-cracks predominantly aligned with axis $y$, represented by the two red elements in Fig. \ref{fig:verf}a (which is what remains after a crack was closed by up to a million  years of secondary creep, or viscous flow). These cracks cause the vertical permeability in these two elements to increase cca 1000-times compared to undamaged shale, while the Biot coefficient increases up to 1 and the initial strength decreases to 10\% of intact shale. 

Figs. \ref{fig:verf}b,c show how damage and pressure propagate after water injection. For this case, the crack band with high water pressure is seen to propagate straight forward, without branching. Now look at stress variation. Fig. \ref{fig:verf}d shows the stress evolution within in the solid part of the first element above the initial damaged elements. Obviously, the damage during post-peak softening is captured in a stable manner. Finally, consider how the Biot coefficient and permeability vary in one cracked element (the first above the initial damaged elements). Fig. \ref{fig:verf}e contrasts the evolution of Biot coefficient in the transverse direction with its constancy in the forward direction, which agrees with experimental observations.

\begin{figure}[h]
\includegraphics[width=1\linewidth]{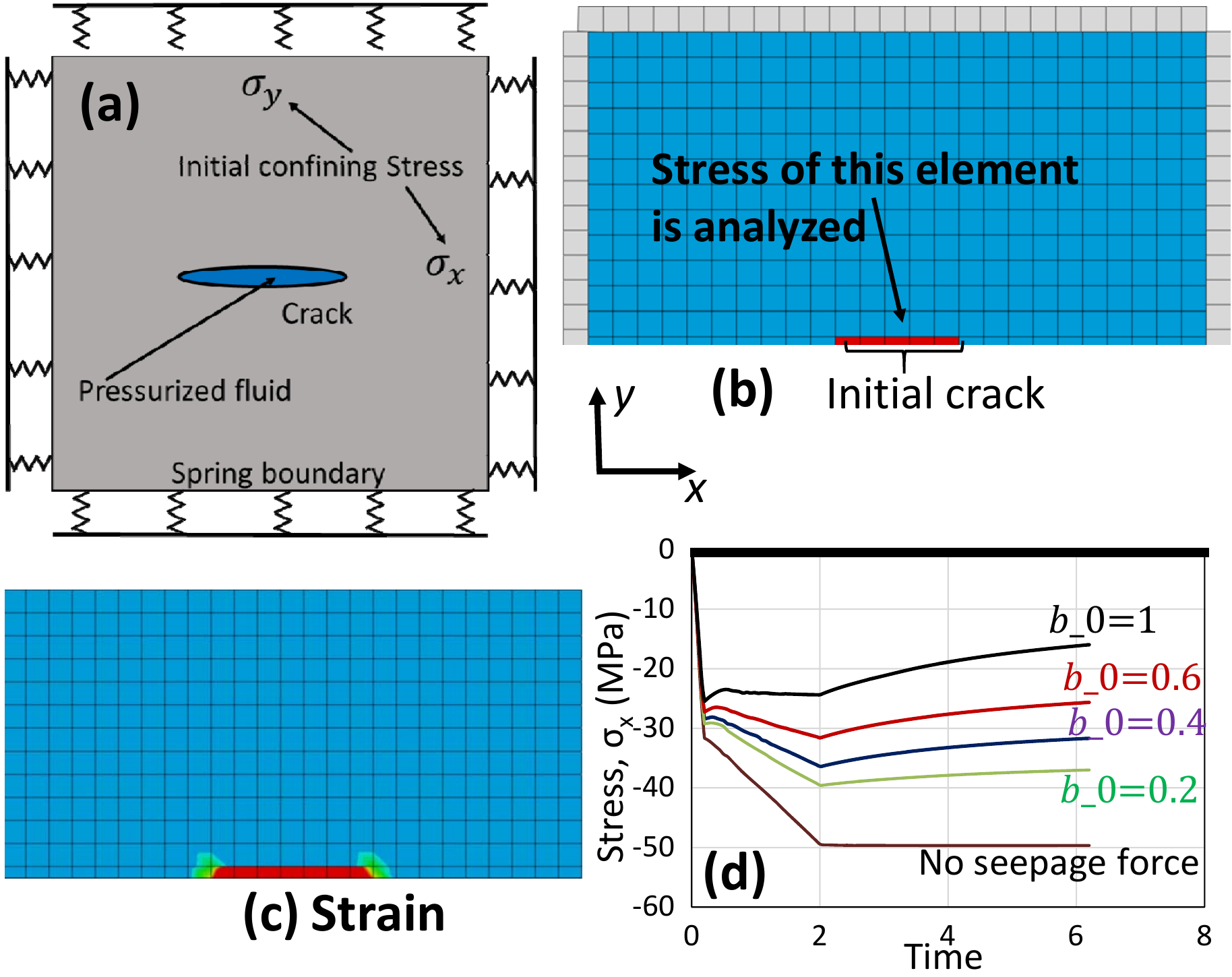}
\caption{a) Pressurized line crack in a 2D domain of a two-phase porous solid (shale) supported by springs at boundaries, subjected to tectonic stresses $T_x$ and $T_y$;  b) FE mesh for one half of the domain; c) Extension of the band of high strain; d) Evolution of stress $\sig_{xx}$ in solid part in the element at the center of initial crack face.    \label{fig:line} }
\vspace{-0.4cm}
\end{figure}

\subsubsection*{Do they seepage forces suffice to induce crack branching?} It is well known in classical fracture mechanics that pressurizing a crack cannot produce tension along the crack faces, and thus cannot initiate lateral crack branching (branching is possible only at the tip of a crack propagating at nearly the Raleigh wave speed). In a preceding study \cite{Chau16}, it was surmised, under various simplifications, that the seepage forces (Eq. \ref{fseep}) would suffice for produce tension along the crack face and thus initiate lateral crack branching. Let us examine it more rigorously. Consider again a horizontal 2D square domain 2.5 m $\times$ 2.5 m, containing one line crack (Fig. \ref{fig:line}a). By virtue of symmetry, only a half-domain is simulated (Fig. \ref{fig:line}b). The water pressure in the line crack is gradually ramped up to reach the maximum of 50 MPa. Water diffusion from the pressurized crack into the shale is simulated via Darcy law. 
 % omit new paragraph?
First we neglect the increase of Biot coefficient due to damage ($\beta=0)$.  Fig. \ref{fig:line}c shows that the damage, as well as the crack, propagates only in the direct extension of the initial line crack, i.e, there is no branching. Fig. \ref{fig:line}d shows the evolution of stress in the solid part, $\sig_{xx}$, along the crack face. The results show that the Biot coefficient can have a major effect and cannot be ignored. 

Lateral crack branching would happen if the stress in the solid phase became positive (tensile) and attained the tensile strength of shale. The results show that this cannot happen, regardless of the tectonic stress value (even if vanishing). Nevertheless, the seepage forces reduce the magnitude of compressive stress along the crack face significantly.

We must thus conclude (as an update of \cite{Chau17}) that the seepage forces alone do not suffice to explain and model lateral branching of hydraulic cracks. So, what other phenomena could explain the lateral branching? Not surprisingly, the explanation is the natural (preexisting) fractures even though they must have been completely closed due to millions of years of secondary creep, or flow. We demonstrate it next.

\subsubsection*{Hydraulic crack branching in two-phase porous solid with closed natural fractures}
\begin{figure}[h]
 \begin{Center}
\includegraphics[width=0.95\linewidth,height=1.6\linewidth]{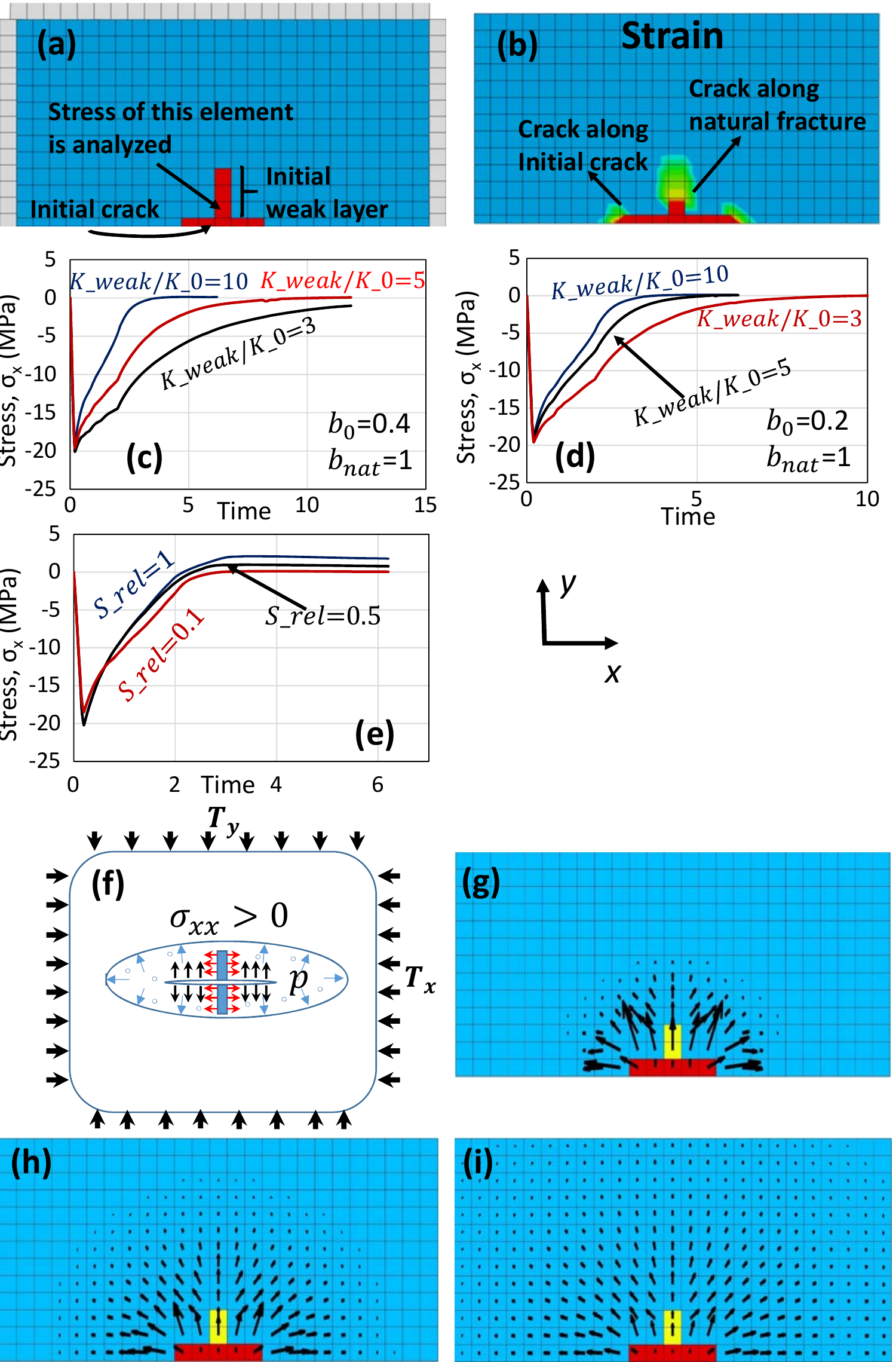}
\caption{a) Schematic line crack with weak layer and spring boundaries; b) Strain and damage evolution; c) stress in solid part of one element in weak layer considering a shale with $b_{0}=0.4$ and different permeabilities for the weak layer; d) stress in solid part of one element in weak layer considering $b_{0}=0.2$ and different permeabilities for the weak layer; e) stress in solid part of one element in weak layer considering $b_{0}=0.4$ and different relative strength values for weak layer; f) schematic of seepage forces; and g-i) evolution of seepage forces  \label{fig:LineF}}
\vspace{-0.4cm}
\end{Center}
  \end{figure}

     %In the previous section, the possibility of branching in a homogeneous two-phase medium was studied. It has been shown that branching is not possible in a homogeneous medium unless due to dynamic crack growth or instability. In this section, the possibility of crack branching in a two-phase medium with weak layers will be studied. In reality, these weak layers always present due to natural fractures formed during the lifetime of shale (million years). Recently, Chau et al. (2018) showed the natural cracks should be closed due to primary and secondary creep caused by overburden pressure. Therefore, in this study, first, we consider these natural fractures are closed, but they are weaker. Especially, we consider higher Biot coefficient and lower strength for these weak layers. Also, in order to study the effect of permeability of these layers on final results, several permeability values will be considered for these weak layers.
In Fig. \ref{fig:LineF}a we now consider the same 2D domain of two-phase porous solid as before, except that now there are two natural weak layers (or preexisting damage bands) in both $x$ and $y$ directions. The crack is uniformly pressurized and water diffuses out. The transverse Biot coefficient within the weak layers that represent the closed natural fractures is $b_{nat} = 1$ because the weak layer (or natural fracture) may consist of separate original crack faces in contact (uncemented by limestone deposit),  while in the intact shale the $b_{ij}$-values increase according to Eq. (\ref{bij}) from the initial value $b_0 = 0.4$ (Fig. \ref{fig:LineF}d) or $0.2$ (Fig. \ref{fig:LineF}d).

Fig. \ref{fig:LineF}b reveals that the hydraulic crack tends to propagate simultaneously along the initial crack and along the weak layer. This confirms that branching can occur if transverse weak layers exist. Further, consider the normal stress parallel to the crack in one element of the weak layer. If this stress attains the tensile strength, a lateral crack branch can initiate and shale branching can happen. Fig. \ref{fig:LineF}b shows the spreading of high and lower transverse strains along both weak layers for the case of Biot coefficient $b=0.4$, with permeability $K_{weak}$ along the weak band 5-times bigger than $K_0$ for intact shale. 

The computed effect of ratio $K_{weak}/K_0$ on the $\sig_{xx}$ evolution in the first element of the weak layer above initial crack is plotted in Fig. \ref{fig:LineF}c,d for the initial Biot coefficients, $b_0$ = 0.4 and 0.2. As water diffuses into the shale, the stress in the weak layer increases from negative to tensile values until it finally reaches the tensile strength of the weak layer. Evidently, a greater difference in Biot coefficient between the weak layer and the shale facilitates, and speeds up the crack branching. 

Finally, to clarify the effect of the transverse tensile strength of the weak layer, three relative strength $S_{rel}$ values are considered in Fig. \ref{fig:LineF}e (here $S_{rel}$ is the damaged-to-intact strength ratio of shale). As seen, a smaller $S_{ref}$ leads to smaller stress, but generally, the effect of $S_{rel}$ is almost negligible. Hence, whether or not the natural cracks are cemented by limestone is almost irrelevant.

It is instructive to see the evolution of the seepage force vectors acting on the mesh nodes, as portrayed in Fig. \ref{fig:LineF}. Fig. \ref{fig:LineF}f shows schematically the seepage forces acting on an ellipse around the crack. Fig. \ref{fig:LineF}g-i illustrates the evolution of seepage forces. Their orientations make it intuitively clear that they must produce in the porous solid a biaxial tension. 
 
From all these observations, it transpires that a major stimulus for crack branching is the difference in the Biot coefficient and in the permeability between the weak layers and the intact shale, as well as the shale mass heterogeneity due to the alternation of weak layers and intact porous solid.

It is worth mentioning that the expansion of solid due to the effect of Biot coefficient has been thought to prevent any tension parallel to the crack face, and thus cause the closing of any lateral crack. The preceding results show that this skeptical view does not extend to a heterogeneous shale mass containing weak layers alternating with intact shale. 
 %11/19*** Add here an explanation why an argument like Detournay's is incorrect.

%  \begin{figure}
%\includegraphics[width=1\linewidth]{Fig/Vecfieldx}
%\caption{Vectors of seepage forces; a) schematic of seepage forces, and b-d) %evolution of seepage forces.   \label{fig:Vecf}}
%  \end{figure}

%\subsubsection*{Hydraulic crack branching in shale with an array of closed nature fractures}

%Consider the horizontal horizontal section of shale with four weak layers (Fig. \ref{fig:oneInj}a). % = Fig. 5
%Water is injecting from one injection point with constant flow rate. Weak layers have $b_{weak}=1,$$S_{rel}=0.1$ and $K_{weak}/K_{0}=100$ (see Fig\ref{fig:oneInj}b). Fig \ref{fig:oneInj}c to Fig\ref{fig:oneInj}e illustrate the evolution of water pressure inside the considered shale. As it can be seen, water initially flows in direction of previously fractured elements next to injection point. The flow into weak layers initially is just due to diffusion, but later when water pressure in these layers increases, crack branching occurs and water flows into weak layer both due to diffusion and Poiseuille flow. This simple example illustrates the significant effect of weak layers on water flow and damage propagation. It clearly shows, the effect of weak layers is not negligible and all currently used models in the industry which neglect the effect of these layers on crack branching are not reliable.

\begin{figure}[h]
\begin{Center}
\includegraphics[width=0.85\linewidth]{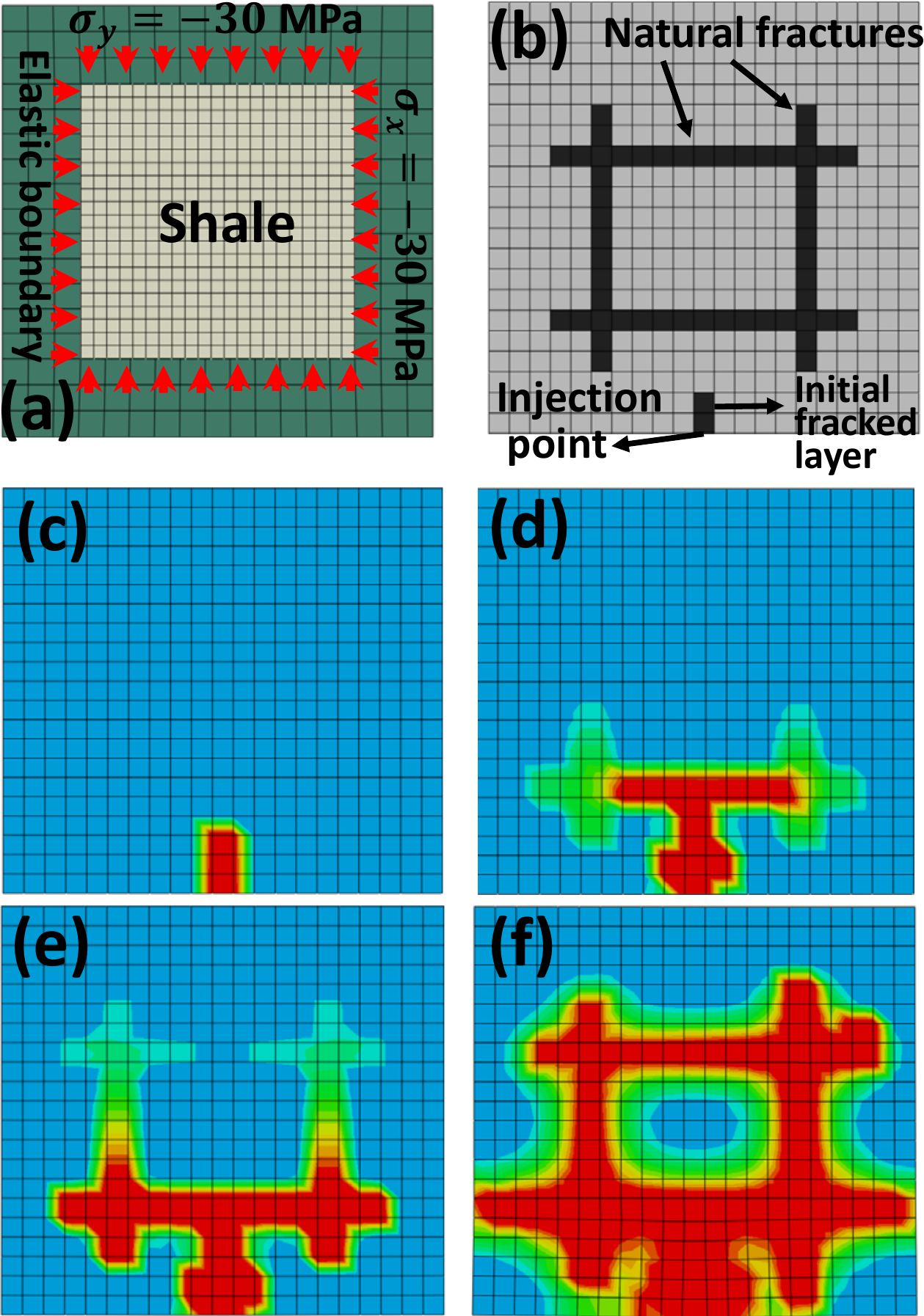}

\caption{FE simulation of hydraulic crack branching in a small domain of shale with several orthogonal weak layers; (a)  Ring of elastic elements providing elastic support of the boundaries; b) FE mesh, preexising natural weak layers, and fracking water inlet; c-f) evolution
of pressure in a shale with weak layers.\label{fig:oneInj}}
\vspace{-0.4cm}
\end{Center}
 \end{figure}

\begin{figure}[h] 
\begin{center}
\includegraphics[width=0.85\linewidth]{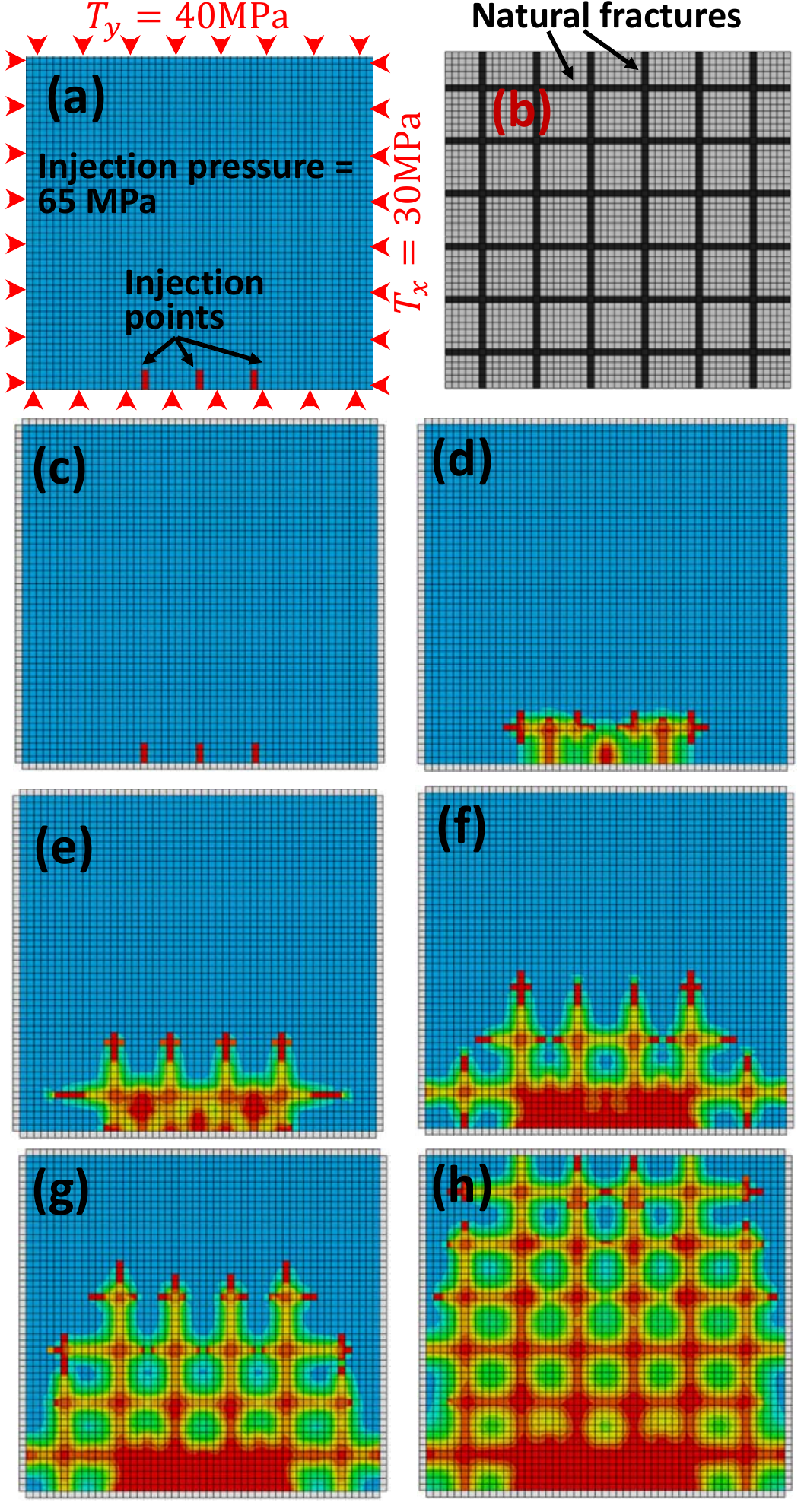}
\caption{2D FE simulation of fracking process in a horizontal domain with a larger system of natural fractures or weak layers (the red zone shows the propagation of high water pressure) \label{fig:Pthree}}
\vspace{-0.4cm}
\end{center}
\end{figure}

To demonstrate the present theory on a larger scale, consider a bigger horizontal section of shale, a square domain 5 m $\times$ 5 m, containing a uniform orthogonal system of closed natural fractures with aligned preexisting weak layers (Fig. \ref{fig:Pthree}). To be more realistic, unequal tectonic stresses are considered in $x$ and $y$ directions; $T_x=$30MPa and $T_y$=40MPa.

Water is injected at three points at the bottom of the figure. Figs. \ref{fig:Pthree}c--\ref{fig:Pthree}h show the evolution of water pressure. Water flow and damage strain are seen to follow the path of weak layers. Extensive branching occurs. Obviously, this branching can create closely spaced hydraulic cracks and thus increase the overall permeability of shale stratum by orders of magnitude, compared to non-branching cracks in intact shale.

\begin{figure}[h]
\begin{center}
\includegraphics[width=0.96\linewidth]{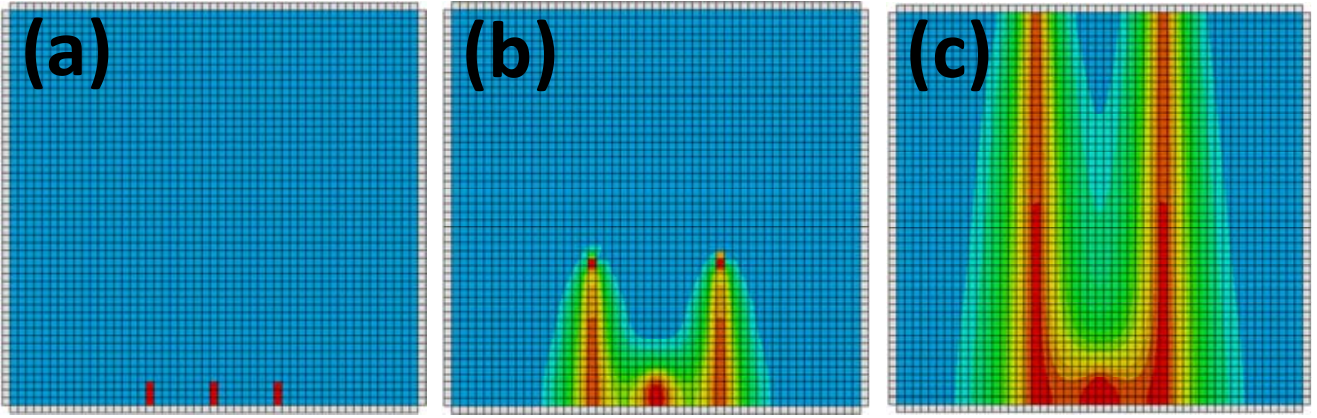}
\caption{Water pressure propagation for the case with no weak layers, all else the same. \label{fig:Now}}
\end{center}
\vspace{-0.4cm}
\end{figure}

It has also been checked that omitting the natural fractures leads to no branching. This is evident from the pressure propagation pattern in Fig. \ref{fig:Now}. This figure also documents the localization instability of parallel crack system \cite{Baz85} (aka the stress shadow effect), which causes that the crack emanating from the middle injection point cannot grow long (the long simultaneous growth of both remaining cracks is made possible by the proximity of the boundaries).

% \begin{figure}  
%\includegraphics[width=0.9\linewidth]{Fig/Schem2}
%\caption{2D FE simulation of fracking process in a horizontal domain with a larger system of natural fractures or weak layers (the red zone shows the propagation of high water pressure). \label{fig:Threeinj} }
% \end{figure}

\subsubsection*{Conclusions}

\begin{enumerate} \setlength{\itemsep}{-1.7mm}

\item The natural fractures have a major effect on hydraulic fracturing and are crucial for its success (although they are currently neglected by the commercial fracking softwares).

%\item The behavior with and without considering the effect of natural fractures were so different that it seems neglecting their effects makes model completely wrong. Unfortunately, all currently used programs are neglecting the effect of weak layers on cracking pattern which makes them unreliable.

\item Even though the natural fractures must have been closed by millions of years of creep, or sealed by mineral deposits, a weak layer of nano- and micro-cracks along these fractures must be expected to facilitate water diffusion.

\item Poromechanics with Biot coefficient depending on the damage of the solid phase must be used in the analysis of fracking. 
    
\item Increase of the Biot coefficient in the transverse direction, caused by oriented cracking damage inflicted by fracking, is essential to achieve crack branching.
    
\item The typical spacing between natural fractures is roughly 0.1 m. This value matches the spacing of hydraulic cracks that is necessary to explain the typical gas production rate at the wellhead.

\item The widespread opinion that preexisting natural fractures somehow explain why the overall permeability of shale mass, inferred from the gas production rate, appears to be about 10,000 times higher than what is measured on shale cores in the laboratory,  has been basically correct. But these fractures are completely closed, do not convey any gas and their role is indirect.

\item a) {\em No porosity $\Rightarrow$ no branching}.
% (mind: granite) 
b) {\em  No seepage forces $\Rightarrow$ no branching}. 
c) {\em  No weak layers $\Rightarrow$ no branching.} 
d) {\em  Constant Biot coefficient $\Rightarrow$ no branching.}  
(Note: consequently, branching in granite is impossible)
    
\end{enumerate}

\acknow{The work at Northwestern was funded by LANL grant to Northwestern University. The last author thanks Emmanuel Detournay, University of Minnesota, for valuable discussions and a stimulating challenge that provoked part of this study.}

\showacknow{} % Display the acknowledgments section

\bibliography{Ref}

%\bibitem{Ter22} Terzaghi, K. (1922). ``Der Grundbruch an Stauwerken und seine Verh\" utung." {\em Die Wasserkraft} (Munich), p. 445.
%\bibitem{Baz40} Ba\v zant, Jr., Z. (1940). "Foundation breakdown behind sheetpiling wall." {\em Technick\' y Obzor} (Prague, in Czech), p. 132.
%\bibitem{Baz53} Ba\v zant, Jr., Z. (1953). {\em Stability of Cohesionless Soil in Curvilinear Upward Seepage.} (in Czech) SNTL, Prague (161 pp.).
%\bibitem{Baz52} Ba\v zant, Jr., Z. (1952) "Graphical solution of foundation equilibrium under weir loaded by seepage drag." (in Czech) {Symp., $70^{th}$ Birthday of Prof. Z. Ba\v zant}. p. 31.
%\bibitem{Ter43}. Terzaghi, K. (1943). {\em Theoretical Soil Mechanics}. J. Wiley, p. 257.

\end{document}